\documentclass[journal,12pt,onecolumn,draftclsnofoot,]{IEEEtran}
\usepackage{algorithm} 
\usepackage{algpseudocode}
\usepackage{amsmath,amsfonts,amssymb}
\usepackage{amsthm}
\usepackage{caption}
\usepackage{cite}
\usepackage{graphicx}
\usepackage{textcomp}

\newtheorem{mydef}{Definition}

\def\BibTeX{{\rm B\kern-.05em{\sc i\kern-.025em b}\kern-.08em
    T\kern-.1667em\lower.7ex\hbox{E}\kern-.125emX}}

\begin{document}

\title{Contract-based Scheduling of URLLC Packets in Incumbent EMBB Traffic}

\author{
	\IEEEauthorblockN{Aunas~Manzoor,
		S. M. Ahsan Kazmi,
		Shashi Raj Pandey
		and~Choong~Seon~Hong,~\IEEEmembership{Senior~Member,~IEEE}}
	\thanks{A. Manzoor, Shashi Raj Pandey and C. S. Hong  are with the Department of Computer Science and Engineering, Kyung Hee University, Yongin 446-701, South Korea (e-mail: \{aunasmanzoor; latif; munir; cshong\}@khu.ac.kr).}
	\thanks{S. M. Ahsan Kazmi is with Network, Cyber and Information Security (NCIS) Lab, Secure System and Network Engineering, Innopolis University, Tatarstan 420500, Russia, and also with the Department of Computer Science and Engineering, Kyung Hee University, Yongin 446-701, South Korea (e-mail: a.kazmi@innopolis.ru).}
}

\maketitle
\begin{abstract}
    Recently, the coexistence of ultra-reliable and low-latency communication (URLLC) and enhanced mobile broadband (eMBB) services on the same licensed spectrum has gained a lot of attention from both academia and industry. However, the coexistence of these services is not trivial due to the diverse multiple access protocols, contrasting frame distributions in the existing network, and the distinct quality of service requirements posed by these services. Therefore, such coexistence drives towards a challenging resource scheduling problem. To address this problem, in this paper, we first investigate the possibilities of scheduling URLLC packets in incumbent eMBB traffic. In this regard, we formulate an optimization problem for coexistence by dynamically adopting a superposition or puncturing scheme. In particular, the aim is to provide spectrum access to the URLLC users while reducing the intervention on incumbent eMBB users. Next, we apply the one-to-one matching game to find stable URLLC-eMBB pairs that can coexist on the same spectrum. Then, we apply the contract theory framework to design contracts for URLLC users to adopt the superposition scheme. Simulation results reveal that the proposed contract-based scheduling scheme achieves up to 63\% of the eMBB rate for the "No URLLC" case compared to the "Puncturing" scheme. 
\end{abstract}
	
\begin{IEEEkeywords}
	 Contract theory, enhance mobile broadband (eMBB), fifth-generation new radio, matching theory, scheduling, ultra-reliable and low-latency communication (URLLC).
\end{IEEEkeywords}

\section{Introduction}
\label{sec:introduction}

As a result of rapid technological developments in the recent years, there has been a growing interest in tactile internet applications such as industrial automation, autonomous vehicles, massive IoT connectivity, and digital entertainment expansion. These novel applications have very stringent and diverse communication requirements such as coverage, data rate, latency and reliability. To meet these diverse communication requirements of diverse applications, the International Telecommunication Union (ITU) has classified fifth-generation new-radio (5G NR) services into three categories: \emph{ultra-reliable and low-latency communication (URLLC)}, \emph{enhanced mobile broadband (eMBB)}, and massive machine type communication (mMTC) \cite{first}. Among these services, URLLC is designed for the event-driven, mission-critical, and industrial scenarios, in which it can contribute to meet the quality-of-service (QoS) requirements such as ultra-low latency and ultra-high reliability. Furthermore, the standard URLLC imposes strict latency and reliability requirements, typically of 1~ms/packet and up to 99.999\% successful packet delivery, respectively \cite{qc,marco2018key}. 
	

Since both the eMBB and URLLC are essential components of communication traffic in 5G cellular networks, various studies have addressed the coexistence issue of these services \cite{matchabedin}. From the perspective of network throughput, the eMBB generates an enormous amount of data communication traffic over cellular networks. Unlike the eMBB, URLLC produces less data as it has stringent latency and reliability requirements. Therefore, the coexistence of these two services involves the challenge of achieving sufficient eMBB throughput while satisfying URLLC requirements \cite{urllc2}. Due to the time-sensitivity of mission-critical applications, such as UAV automation, autonomous vehicular control, and critical medical apparatus management, URLLC is prioritized over eMBB for scheduling. In general, eMBB scheduling involves the enhancement of network throughput to improve spectral efficiency while the reliability of packet delivery is ensured through re-transmissions. However, eMBB scheduling approaches may not ensure the reliability and latency thresholds required for URLLC, and thus cannot be applied in URLLC scheduling. On the contrary, URLLC involves the transmission of short packets within certain latency and reliability bounds.
	

One well-thought way to handle the aforementioned constraints is using re-transmissions approach. However, it is impractical to simultaneously satisfy the contradicting requirements of latency and reliability in URLLC only through re-transmissions approach. For instance, a greater number of transmissions ensures reliability while compromising latency, and vice versa. To address this challenge, small-packet communication has been proposed for URLLC, as it can meet the reliability requirements at the cost of reduced spectrum efficiency due to additional control overhead. However, it does not suffice as an efficient scheduling scheme is further required for the coexistence of URLLC and eMBB services such that the penalization of the eMBB traffic is minimized \cite{popovski2018wireless}. 
	
In this regard, conventional scheduling schemes based on orthogonal channel allocation are proven to be inefficient and underutilized when applied to the coexistence scenarios of URLLC and eMBB networks. In fact, the third generation partnership project (3GPP) suggested a short and long transmission time interval (TTI)-based frame distribution for these coexistence scenarios. In this frame distribution, eMBB traffic is scheduled for a long TTI and URLLC is opportunistically scheduled for a short TTI over the existing eMBB traffic by adopting a puncturing or superposition scheme \cite{test}. The puncturing scheme involves the scheduling of URLLC packets by halting the eMBB communication during the URLLC transmission for the duration of a short TTI. Note that the puncturing scheme can significantly reduce the throughput of eMBB users. Thus, to compensate for this loss, the superposition scheme is  proposed that involves the non-orthogonal scheduling of both eMBB and URLLC traffic on the single channel simultaneously. This is achieved by exploiting the non-orthogonal multiple access (NOMA) scheme, in which the difference between the channel gains of the cellular users is exploited to pack the coexisting users on a single channel resource \cite{kazmi2018coordinated,8406264}. 
	
	 
	\subsection{Contribution}
	In this study, we propose an efficient scheduling scheme for the coexistence of eMBB and URLLC by dynamically adopting puncturing or superposition schemes. For this purpose, we consider a cellular network in which many URLLC and eMBB users are associated with a base-station (BS). The BS performs eMBB scheduling at the start of a long TTI, and URLLC scheduling is performed for a short TTI using the puncturing or superposition scheme. To meet the latency requirements of URLLC users, their scheduling is performed in the same or next URLLC TTI of the scheduling request. Moreover, the reliability requirement is met by either using the puncturing or the superposition scheme to ensure certain channel quality. 
	To perform the pairing between URLLC and eMBB users, we apply a one-to-one matching scheme that is based on the preference profiles of the eMBB and URLLC users. The BS attempts to enhance the eMBB throughput by reducing the effects of URLLC scheduling, which is achieved by selecting the superposition scheme. To encourage URLLC users to opt for the superposition scheme, we propose a contract-based incentive mechanism that involves sharing the payoff received by the BS for applying the superposition scheme with the contributing URLLC users. In addition, the URLLC users prefer opting for the superposition scheme unless their QoS requirements are satisfied; otherwise, they select the puncturing scheme.  
	
	In the case of the proposed contract-based incentive mechanism, superposition is well-suited for both URLLC and eMBB users. However, in some cases, both type of the users experience similar channel gains that makes the adoption of the superposition scheme infeasible. In these cases, the puncturing scheme is essential. Therefore, we propose applying the puncturing or superposition scheme based on network and channel conditions. For this purpose, we use the contract theory framework \cite{bolton2005contract} to design a bundle of contracts by the BS for URLLC users. Contract theory is useful because the BS cannot reveal complete information (i.e., channel gain, willingness for superposition, and the matching with a particular eMBB user) from URLLC users in a timely manner due to the strict latency requirements. As a result, an asymmetric information problem arises due to the lack of complete information that is solved by using contract theory \cite{laffont2009theory}. Note that the pairing of eMBB and URLLC users before designing the contracts using the one-to-one matching is essential for better spectrum efficiency. In case of not using the matching pairs, it is possible to select the pairs which may not coexist on a single channel.

	The main contributions of this paper are summarized as follows:
	
	\begin{itemize}
		\item We model the problem of URLLC and eMBB coexistence, in which eMBB users are modeled using the Shannon rate, and finite block length codes are used to model the rates for URLLC users. In addition, we model the superposition and puncturing framework for the coexistence of URLLC and eMBB users on the same channels. 
		\item For URLLC scheduling, we formulate the optimization problem to maximize the eMBB rate under the URLLC QoS requirements of latency and reliability to optimize the puncturing or superposition scheme and URLLC power allocation.  
		\item To solve the formulated problem, we first pair each URLLC user with a suitable eMBB user. This pairing is performed by applying the one-to-one matching considering the preference profiles of both participants.  
		\item Based on the pairing, appropriate contracts are designed for each URLLC type. The URLLC type refers to the classification of URLLC users for the adoption of superposition scheme. After verifying the feasibility and optimality of the contracts, power allocation to the URLLC users is performed according to their utility constraints. Furthermore, we formulate a convex problem that maximizes the BS profit by optimizing the power allocation.  
		\item Numerical results validate the performance of the proposed contract-based scheme. The results demonstrate that the proposed contract-based superposition scheme achieves up to 63\% of the eMBB rate for the non-URLLC case compared to the puncturing scheme. 
	\end{itemize}
	
 The remainder of this paper is organized as follows. The related work is summarized in Section \ref{secRW}. The system model is presented in Section \ref{secsys}. Subsequently, in Section \ref{secProb}, the problem is formulated to maximize the eMBB rate subject to URLLC requirements. In Section \ref{secContMatch}, after performing the one-to-one matching, a contract design for URLLC users is presented which is used for the resource allocation to the URLLC users. Finally, numerical results and conclusions are provided in Sections \ref{secSim} and \ref{secCon}, respectively.   
	

\section{Related Works}
\label{secRW}
In this section, we discuss some of the significant related works and challenges, which are grouped into three categories: (a) 5G-NR, (b) contract theory, and (c) matching theory. 

\subsection{5G-NR}

Recently, numerous puncturing-based scheduling schemes have been proposed in the literature.
For instance, the authors in \cite{anand} utilized puncturing and superposition schemes to schedule URLLC traffic over pre-scheduled eMBB communication. However, the authors did not consider the reliability constraints of URLLC communication. A statistical analysis of URLLC communication over a wireless channel was performed in \cite{angjelichinoski2018statistical}. The authors proposed a method of selecting a transmission rate according to the channel conditions and reliability requirements. 
In \cite{kassab2018coexistence}, the authors considered a CRAN environment where the decoding of eMBB traffic was performed on the cloud while URLLC decoding was performed by edge nodes to meet the latency requirements. In \cite{urllc1}, the authors solved the URLLC resource allocation problem in the short blocklength regime. However, the global optimal solution was identified in the subset of the feasible region only. The authors demonstrated the insignificance of power control for small URLLC packets. In \cite{urllc3d2d}, the authors proposed URLLC packet transmission among device-to-device (D2D) users, in which the D2D pairs communicated opportunistically for short packets. However, the authors did not consider the reliability requirements in their formulation. In \cite{urllcmachinelearning}, machine learning based adaptive TTI interval is proposed for the scheduling in eMBB and URLLC coexistence networks. 

	  
	 The problem of ensuring ultra-low latency and ultra-reliability has been addressed in the literature. For instance, backbone network latency can be improved using a dedicated link for URLLC communication. Similarly, fronthaul latency reduction is possible by reducing the transmission overhead. Furthermore, the control signaling mechanism can be improved to eliminate the signaling latency in the LTE systems \cite{hyoungju}. In \cite{madyan}, the authors proposed a risk-sensitive based formulation for the coexistence problem of eMBB and URLLC traffics that aims at maximizing the eMBB data rate while considering the URLLC reliability. In \cite{shashi}, the authors proposed a scheduling scheme for the URLLC downlink traffic.  
	 The main cause of reliability losses in current LTE systems is erroneous channel estimation. Therefore, URLLC reliability can be increased by the improving channel estimation, which can be achieved by improving the control signaling mechanism. 
	 As mentioned earlier, one solution to the problem of simultaneously meeting reliability and latency requirements involves reducing the packet size in URLLC, which results in meeting the reliability constraints for a given latency at the cost of low achievable rates. 
	 Furthermore, spatial diversity can be used to achieve an improved URLLC reliability i.e., using multiple transmitters for sending duplicate URLLC packets. In this way, the required reliability can be achieved at the cost of spectrum efficiency \cite{spadiv}.  

\subsection{Contract Theory}

Contract theory has been widely used in various wireless communication schemes for situations involving information asymmetry, as well as to encourage agents to contribute to tasks assigned by the principal \cite{zhang2017survey}. For instance, the study conducted in \cite{zhang2015contract} proposed an incentive mechanism to encourage D2D users to share content. Similarly, in a cloud radio access network, contract theory was used to motivate the content providers to rent the cache to the network operator \cite{le2019joint}. In addition, the use of contract theory for the case of incomplete information was exploited in \cite{duan2012cooperative}. In \cite{contract1blockchain}, the authors addressed the problem of secure data sharing in the Internet of Vehicles by leveraging blockchain. Moreover, to address the problem of minor selection, the authors applied contract theory. In \cite{contract2}, the authors proposed task offloading from the BS to nearby underutilized vehicular fog nodes. They proposed a contract-matching based incentive and task assignment scheme. In \cite{contractd2d}, the contract theory was used to model communication among D2D users. A multi-principal multi-agent problem was mapped to the D2D communication problem. The aforementioned works thus indicate the utility of using contract theory in real-world problems. 

\subsection{Matching Theory}

In economics, a Nobel prize-winning mathematical framework called matching theory has been developed that is applied in the formation of collectively valuable groups among participants. Recently matching theory has been extensively used for efficient resource management in wireless networks \cite{matchmag}. For instance, the authors of \cite{matching1} used matching theory to associate users for task offloading in mobile edge computing (MEC). Similarly, the problem of user association and resource allocation in a fog network was solved using matching theory for the two-way association between fog nodes and (IoT) users \cite{matchabedin}. Furthermore, a matching-based D2D resource allocation with interference management was proposed in \cite{matchkazmi}. The aforementioned research that demonstrates the significance and contribution of matching-based approaches in wireless networks. 

To the best of our knowledge, the present study is the first to employ contract theory to address the problem of coexistence of URLLC and eMBB users in cellular networks.
	
\section{System Model and Problem Formulation}
\label{secsys}

	\begin{figure}[!t]
		\centering			
		\includegraphics[width=0.5\linewidth]{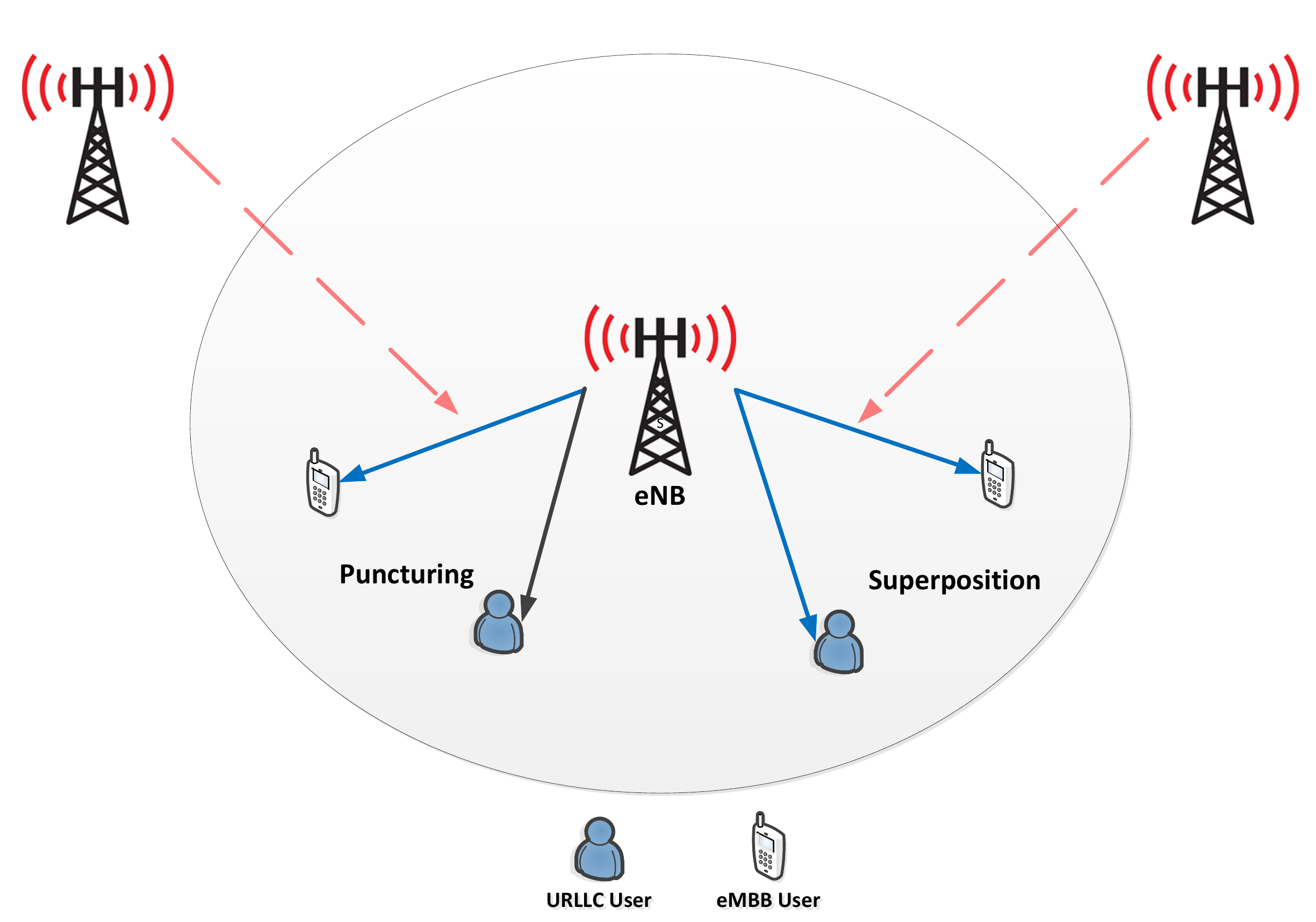}
		\setlength{\belowcaptionskip}{-8pt}
		\caption{System model for the coexisting URLLC and eMBB networks.}
		\label{sys}
	\end{figure}

\begin{figure}[t]
	\centering			
	\includegraphics[width=0.5\linewidth]{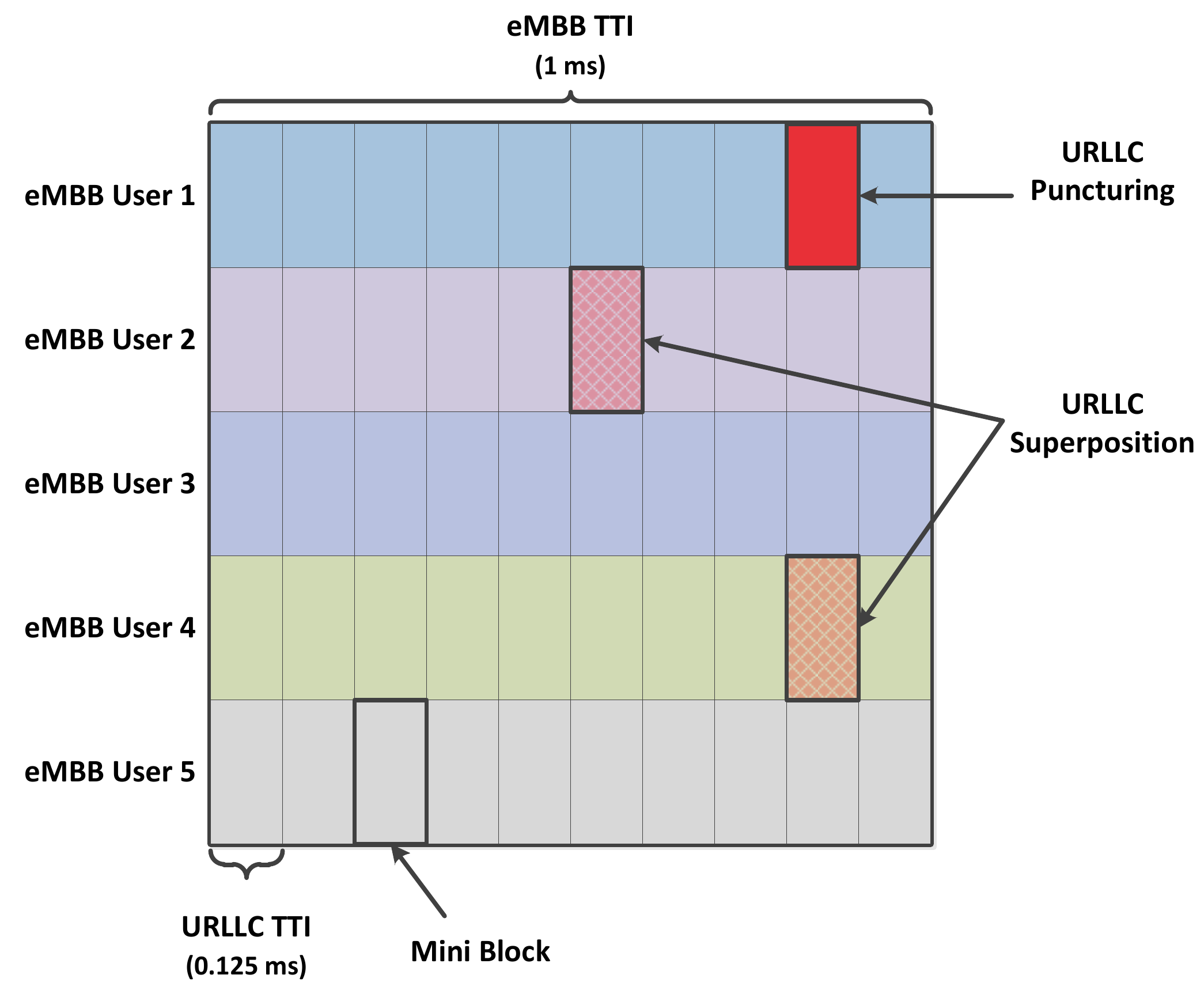}
	\caption{Resource block frame structure for eMBB and URLLC scheduling.}
	\label{frame}
\end{figure}

In the system model, we consider a downlink radio access network (RAN) consisting of a set of BSs denoted by $\mathcal{J} = \{1,2,\cdots,J \}$, each connected to the core network through backhaul links. Each BS $j \in \mathcal{J}$  is associated with the corresponding sets of URLLC and eMBB users denoted by $\mathcal{U} = \{1,2,\cdots,U\}$ and $\mathcal{E} = \{1,2,\cdots,E\}$, respectively as shown in Figure \ref{sys}. The eMBB users are scheduled according to the standard LTE scheduling while URLLC communication is overlaid on the pre-scheduled eMBB traffic. Specifically, both eMBB and URLLC networks coexist to share the same spectral resources consisting of a set of resource blocks (RBs) denoted by $\mathcal{K} = \{1,2,\cdots,K\}$. 
In order to enable coexistence, either the puncturing scheme or the superposition scheme can be used.
Figure \ref{frame} demonstrates the frame structure for eMBB and URLLC scheduling through both the aforementioned schemes. Each eMBB user $e \in \mathcal{E}$ is assigned a timeslot of duration $1$~ms and bandwidth $f$ of an RB $k$. We assume eMBB users are pre-scheduled and each eMBB user experiences a level of signal-to-interference-plus-noise ratio (SINR) that is known to the BS. The arrival of a URLLC transmission request is modeled as a Poisson distribution with an arrival rate of $\lambda$. In each mini-slot of duration $0.125$~ms, there is a random arrival of URLLC users that are scheduled over the incumbent eMBB allocation using the puncturing or superposition scheme.

\subsection{Wireless Model}
In our model, we assume a saturated network scenario in which eMBB users always have packets to transmit and the number of eMBB users are more than the available BS resources. Then, the BS assigns its resources to a set of optimal eMBB users for downlink communication. On the other hand, with respect to the URLLC traffic, we model the arrival of a URLLC request using Poisson distribution \cite{anand2018resource}.

\subsubsection{eMBB traffic}
In wireless networks, we have orthogonal and non-orthogonal channel access schemes. In the scenarios of URLLC and eMBB coexistence, the orthogonal channel access scheme corresponds to the puncturing scheme and non-orthogonal channel access scheme corresponds to superposition scheme, respectively. Upon the arrival of URLLC requests, appropriate mini-slots are allocated to URLLC users according to two schemes: 1) puncturing scheme, in which eMBB transmission is stopped to schedule a URLLC user, and 2) superposition scheme, in which both eMBB and URLLC can operate in the same mini-slot while utilizing the successive interference cancellation (SIC) technique. Under the superposition scheme, the SINR at the eMBB user $e \in \mathcal{E}$ from the BS $j$ is defined as:
\begin{equation}
\gamma_{ej}^{sup} =  \frac{P_{ej} g_{ej}}{I_{urllc}+ N_0},
\label{eqGammE1}
\end{equation}
where $P_{ej}$ and $g_{ej}$ are the transmit power and channel gain, respectively, from the BS $j$ to the eMBB user $e \in \mathcal{E}$. $I_{urllc} = P_{uj} g_{uj}$ denote the received interference from the coexisting URLLC users and $N_0$ denote the noise level. Similarly, the SINR for the puncturing scheme denoted by $\gamma_{ej}^{pun}$ from the BS to the eMBB user is expressed as follows: 
\begin{equation}
\gamma_{ej}^{pun} =  \frac{P_{ej} g_{ej}}{N_0}.
\label{eqGammE2}
\end{equation}

Using (\ref{eqGammE1}) and (\ref{eqGammE2}), the corresponding eMBB rate is calculated as follows:

\begin{equation}
R_{ej}\left(\gamma_{ej}^{(\cdot)}\right) = W \log \left(1 + \gamma_{ej}^{(\cdot)}\right), 
\end{equation}
where $W$ denotes the bandwidth of RB, and $\gamma_{ej}^{(\cdot)}$ denotes the corresponding SINR for the superposition or puncturing case. The total eMBB rate using superposition and puncturing scheme is expressed as $R_{ej} = R_{ej}\left(\gamma_{ej}^{(sup)}\right) + R_{ej}\left(\gamma_{ej}^{(pun)}\right)$.

\subsubsection{URLLC traffic}

The traditional Shannon capacity rates cannot be used due to small packet size in URLLC traffic \cite{urllc1}. Therefore, we use the finite blocklength to define the URLLC rate to deliver the number of bits using RB $k$ with a certain error probability $\epsilon_u$. 

The SINR for the puncturing and superposition cases denoted by $\gamma_{uj}^{(\cdot)}$ is given below: 

\begin{equation}
\gamma_{uj}^{sup} =  \frac{P_{uj} g_{uj}}{I_{embb}+ N_0},
\label{eq:sup}
\end{equation}

\begin{equation}
	\gamma_{uj}^{pun} = \frac{ P_{uj} g_{uj}}{N_0},
	\label{eq:pun}
\end{equation}
where $P_{uj}$ and $g_{uj}$ denote the transmit power and channel gain, respectively, from the BS $j$ to the URLLC user $u \in \mathcal{U}$. $I_{embb} = P_{ej} g_{ej}$ denotes the received interference from the coexisting eMBB users.

Using (\ref{eq:sup}) and (\ref{eq:pun}), the corresponding URLLC rate based on finite block length codes is given as follows \cite{popovski20185g}:  

\begin{equation} 
\begin{split}
R_{uj}\left(\gamma_{uj}^{(\cdot)}\right)  & = W \log(1 + \gamma_{uj}^{(\cdot)}) - \\
& \sqrt{\frac{1}{m_u} \left( 1 -\frac{1}{(\gamma_{uj}^{(\cdot)} +1)^2}\right)}\frac{Q^{-1} (\epsilon)}{\ln2},
\end{split}
\end{equation}
where $m_u$ denotes the packet size of URLLC user $u$, $Q^{-1} (\cdot)$ denotes the inverse Q-function and $\epsilon$ denotes the reliability threshold of URLLC, respectively. 

The adoption of superposition or puncturing scheme by the URLLC user $u$ is represented by $x_{uj}^k$ and $z_{uj}^k$, respectively and defined as follows:

\begin{equation*}
x_{uj}^k =
\begin{cases}
1, &\text{if URLLC user $u $ users superposition},\\
0, & \text{otherwise}.
\end{cases}
\end{equation*}

\begin{equation*}
z_{uj}^k =
\begin{cases}
1, &\text{if URLLC user $u $ uses puncturing},\\
0, & \text{otherwise}.
\end{cases}
\end{equation*}


For the coexistence of eMBB and URLLC, we set the eMBB TTI to $1$~ms and the URLLC TTI (also called mini-slot) to $0.125$~ms. In each eMBB TTI, we spatially divided the TTI to form a set $\mathcal{K}$ of RBs, where each RB $k \in \mathcal{K}$ has spectrum bandwidth $f$. At the start of each subframe, eMBB users are scheduled for an eMBB TTI of $1$~ms and spectrum bandwidth $f$.

Fig. \ref{frame} indicates that eMBB users are scheduled for eMBB TTI and URLLC users are scheduled in the mini-slot. Note that an appropriate selection of mini-slots for URLLC traffic is required for spectrum efficiency and to achieve high eMBB rate. Each eMBB user experiences a different SINR level, depending on the distance from the BS and the corresponding channel gain. Therefore, suitable eMBB users can operate on the same channel with an appropriate URLLC user. In a traditional superposition scheme, URLLC users receive interference from coexisting eMBB users, as given in (\ref{eq:sup}), which can compromise reliability. Therefore, we propose a superposition scheme in which URLLC reliability is not affected. 

 For the case of superposition in which $x_{uj}^k = 0$, both eMBB and URLLC transmit on the same channel $k$. The BS $j$ transmits the superposed signal $x_j = P_{ej}x_e + P_{uj}x_u$ to both eMBB and URLLC users, where $x_e$ and $x_u$ denote the messages for eMBB and URLLC users, respectively \cite{noma}. The superimposed signal is received by both eMBB and URLLC users, and each user performs SIC to decode its own message. The eMBB user decodes the interference signal from the URLLC user and cancels it to get its own message. Conversely, the URLLC user can receive the message without any SIC. The corresponding rates for the eMBB and URLLC users are given as follows: 

\begin{equation}
	y_{ej} = R_{ej} \left( \frac{P_{ej} g_{ej} }{ P_{uj} g_{uj} + N_0 }  \right) , 
	\label{supEmbb}
	\end{equation}  
	
	\begin{equation}
	y_{uj} = R_{uj}\left(\frac{P_{uj} g_{uj}}{N_0}\right).
	\label{supUrllc}
	\end{equation}  

\subsection{Problem Formulation}
\label{secProb}

The objective of this work is to maximize the eMBB rate while satisfying the latency and reliability constraints for the URLLC users.  
Therefore, the goal is to select the optimal mini-slots from the pre-scheduled eMBB traffic such that the latency constraints are not violated. Moreover, it is also required to choose the optimal channel and power to satisfy the reliability constraints for the URLLC traffic.

\begin{subequations}\label{eq:obj1}
	\begin{align}
	\max_{\boldsymbol{x}, \boldsymbol{z}, \boldsymbol{p}} \tag{\ref{eq:obj1}} \quad 
	& \sum\limits_{ j \in \mathcal{J}} \sum\limits_{k \in \mathcal{K}} \sum\limits_{e \in \mathcal{E}} \left(T -  t\sum\limits_{u \in \mathcal{U}}z_{uj}^k\right) R_{ej},  \\
	\text{s.t.}\quad &  \label{eq:p1const1}  \sum\limits_{k \in \mathcal{K}} \sum\limits_{j \in \mathcal{J}} \left( x_{uj}^k + z_{uj}^k \right)= \lambda , \quad \forall u \in \mathcal{U}, \\
	&\label{eq:p1const2} \sum\limits_{j \in \mathcal{J}} \sum\limits_{k \in \mathcal{K}} \left( x_{uj}^k + z_{uj}^k \right) R_{uj}\left(\gamma_{uj}^{(\cdot)}\right) \geq \epsilon_u , \quad \forall u \in \mathcal{U}, \\	
	&\label{eq:p1const3} x_{uj}^k, z_{uj}^k \in \{0,1\} , \quad \forall u \in \mathcal{U}, \forall j \in \mathcal{J}, k \in \mathcal{K}, \\
	&\label{eq:p1const4} p_{uj} \geq 0 , \quad \forall u \in \mathcal{U}, j \in \mathcal{J}.
	\end{align}
\end{subequations}

The objective to maximize the network rate of eMBB users is dependent on the superposition and puncturing schemes represented by $x_{uj}^k$ and $z_{uj}^k$, respectively. $P_{uj}$ represents the power allocated to the URLLC users. The constraint in (\ref{eq:p1const1}) ensures that URLLC packets are strictly scheduled just after the arrival of $\lambda$ transmission requests. (\ref{eq:p1const2}) ensures ultra reliability $\epsilon_u$ of each URLLC user $u$. Equations (\ref{eq:p1const3}) and (\ref{eq:p1const4}) contain the bounds for the decision variables. 

The aforementioned problem optimizes the use of superposition/puncturing scheme for URLLC traffic scheduling. The problem is a large mixed integer linear programming (MILP) problem with very high complexity and is difficult to solve. To solve this problem, we use the contract theory framework to determine the willingness of the URLLC users to adopt the superposition scheme. In order to reduce the number of contracts for each URLLC user and to make the contract design scalable, we use one-to-one matching between the eMBB and URLLC users. After the optimal contract design for the superposition, the power allocation is performed for the URLLC users to meet the reliability constraint.

\section{Solution Approach}
\label{secContMatch}
\subsection{Contract Theory for Superposition}     
\label{secCont1}
The wireless spectrum and number of RBs at each BS are limited. In the coexistence environment of URLLC and eMBB, the aim of the BS is to meet the URLLC QoS requirements and maximize the served number of eMBB users. This can be achieved by efficiently packing the URLLC users on the ongoing eMBB communication using the superposition. Therefore, each BS requires an incentive mechanism to encourage URLLC users to opt for superposition. Moreover, the URLLC users must also meet their QoS requirements. However, the BS does not know the willingness of URLLC users in the network to opt for superposition. Therefore, the problem of optimal resource allocation becomes difficult for each BS, which results in information asymmetry between the BS and URLLC users. We use the contract theory framework to motivate URLLC users for superposition, and each BS designs a bundle of contracts for the URLLC users. 

\begin{figure}[t]
	\centering			
	\includegraphics[width=0.5\linewidth]{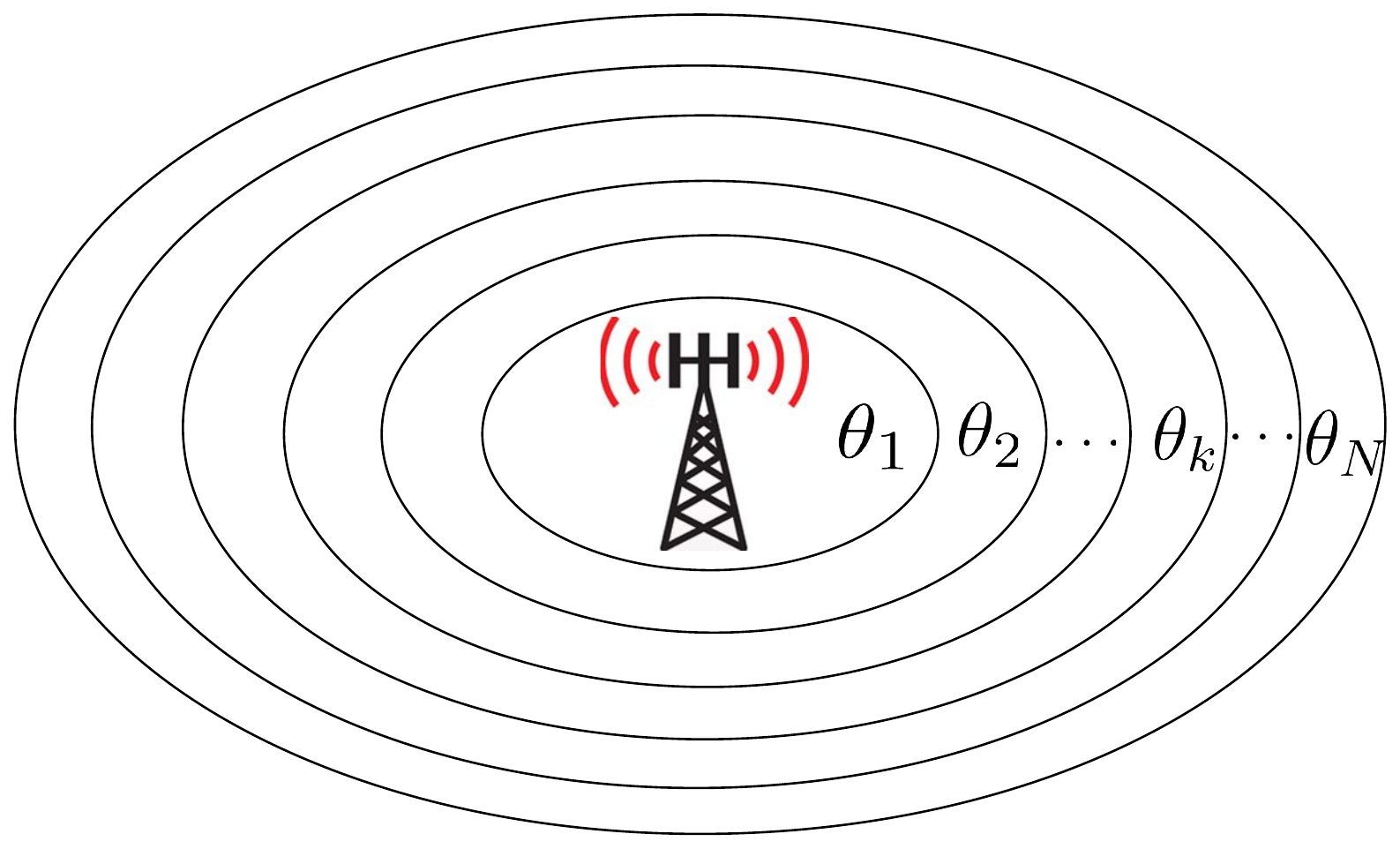}
	\caption{URLLC user tiers of willingness to opt for superposition.}
	\label{type}
\end{figure}

First, we define the types of URLLC users based on their preference to opt for superposition based on the geographical location of the URLLC user, as illustrated in Fig. \ref{type}. A URLLC user with a higher type is more willing to opt for the superposition than a URLLC user with a lower type. 

\begin{mydef}
	We define the willingness of URLLC users to opt for superposition by considering $N$ number of tiers in a geographical cell. The URLLC users closer to the BS are more willing to opt for superposition. Therefore, the URLLC users are classified into a set $\mathcal{\varTheta} = \{\theta_1, \theta_2,\dots,\theta_N\}$ with their corresponding willingness for superposition provided in descending order $\theta_1>\dots>\theta_n>\dots>\theta_N$.
	\label{def1}
\end{mydef}

\subsubsection{Utility of URLLC Users}

The utility of URLLC user $u$ is defined as follows:

\begin{equation}
    U_u = \begin{cases}
     \theta_n y_{uj}  - \beta_u + \varsigma \beta_e, &\text{if $\theta_n \geq \upsilon(\mu(u)) $},\text{ $x_{uj}^k = 1 $},  \\
     \theta_n y_{uj}  - \beta_u, &\text{if $\theta_n < \upsilon(\mu(u))$},\text{ $z_{uj}^k = 1$},
\end{cases}
\label{util_u}
\end{equation}
where $\upsilon(\mu(u))$ denotes the parameter of choosing between superposition or puncturing scheme by the URLLC user which is function of the matched eMBB and URLLC pair $\mu(u)$. Depending on the utility obtained from the matched pair, each URLLC user selects the optimal superposition or puncturing scheme ${x_{uj}^k}^*$ or ${z_{uj}^k}^*$, respectively. $\beta_u$ and $\beta_e$ denote the price per mega bytes (MB) paid by the URLLC and eMBB users to the BS, respectively, and $\varsigma$ denotes the proportion of the incentive paid to the URLLC user on choosing the superposition scheme.

\subsubsection{BS Utility}

The utility of the BS $j$ is the difference between the profit obtained from URLLC and eMBB users, and the resources allocated to them:

\begin{equation}
\begin{split}
U_j & = \xi \left( \sum\limits_{u \in \mathcal{U}} \beta_u + \sum\limits_{e \in \mathcal{E}}  \beta_e \right) - \\
&\zeta \left( \sum\limits_{u \in \mathcal{U}} \pi_u\left(R_{uj}\left(\gamma_{uj}^{(\cdot)}\right)\right) + \sum\limits_{e \in \mathcal{E}}  \pi_e(R_{ej}) \right),
\end{split}
\label{util_bs}
\end{equation}
where $\xi$ and $\zeta$ denote the normalizing constants. We consider $\beta_u >> \beta_e$, which signifies that the price $\beta_u$ paid by URLLC users is significantly greater than the price $\beta_e$ paid by eMBB users. Therefore, the BS preferes scheduling URLLC users over eMBB users. $\pi_u(\cdot)$ and $\pi_e(\cdot)$ denote the expenses of the BS in terms of channel and power allocation to URLLC and eMBB users, respectively.  

Next, we present the one-to-one matching scheme between the URLLC and eMBB users which is essential for the scalability of the designed number of contracts. 
\subsection{Matching URLLC and eMBB users}
\label{subsec:MGF}
In the coexistence scenario of eMBB and URLLC, each URLLC user can be paired with only one eMBB user and vice versa. Therefore, we formulate a one-to-one matching game between URLLC and eMBB networks. The matching is defined as follows:

\begin{mydef}
A \emph{matching} $\mu$ is defined by a function from the set $ \{\mathcal{E} \cup \mathcal{U}\} \text{ into the set of elements of } \{ \mathcal{E} \cup \mathcal{U} \}$ such that: 
\begin{enumerate}
	\item  $\left| {\mu (e)}\right| \le 1 \ $\text{and}$ \ \mu (e) \in \mathcal{U},$
	\item  $ \left| {\mu (u)} \right| \le 1 \ $\text{and}$ \ \mu (u) \in  \mathcal{E}\cup \phi,$ 
	\item  $\mu(e)=u$ if and only if $u$ is in $\mu(e)$,
\end{enumerate}
where $\mu (e)=\{u\} \Leftrightarrow \mu (u)=\{e\}$ for $\forall e \in {\mathcal{E}}, \forall u \in {\mathcal{U}}$ and $\left| {\mu(\cdot)} \right|$ represents the cardinality of the matching outcome $\mu(\cdot)$. 
\end{mydef}
The first two properties ensure that the matching between the eMBB and URLLC networks has a one-to-one relation in that an eMBB user $e$ can be paired with only one URLLC user $u$. Additionally, when a matching pair cannot be paired for any reason, we have $\mu (n)=\phi$.

\begin{algorithm}[t!]
	
	\caption{URLLC-eMBB matching algorithm}
	\label{algo:association}
	
	\begin{algorithmic}[1]
		\State \textbf{\emph{Phase 1: Initialization}}:
		\State \textbf{input}: $\mathcal{P}_u$, $\mathcal{P}_e$, $\forall u, e$
		\State \textbf{initialize}: $t = 0$, $\mu^{(t)} \triangleq {\{\mu(u)^{(t)},\mu(e)^{(t)}\}}_{u\in \mathcal{U}, e \in \mathcal{E}}=\emptyset$, ${\mathcal{R}_e}^{(t)}=\emptyset$, ${\mathcal{P}_u}^{(0)}=\mathcal{P}_u$, ${\mathcal{P}_e}^{(0)}=\mathcal{P}_e$, $\forall u,e$ 
		\State \textbf{\emph{Phase 2: Matching}}: 
		\Repeat 
		\State $t \gets t+1$
		\For {$e \in \mathcal{E}$, propose $u$ according to ${\mathcal{P}_e}^{(t)}$}
		\If {$e \succ_u \mu(u)^{(t)}$}
		\State ${\mu(u)}^{(t)} \gets {\mu(u)}^{(t)}\setminus e'$
		\State ${\mu(u)}^{(t)}\gets e$
		\State ${\mathcal{P'}_u^{(t)}}= \{e'\in {\mu(u)^{(t)}} | e\succ _ue'\}$
		\Else
		\State ${\mathcal{P''}_u^{(t)}}= \{e \in {\mathcal{E}} |\mu(u)^{(t)}\succ _ue\}$
		\EndIf
		\State ${\mathcal{R}_u}^{(t)}= \{{\mathcal{P'}_u^{(t)}}\} \cup \{{\mathcal{P''}_u^{(t)}}\}$
		\For {$l \in {\mathcal{R}_u}^{(t)}$ } 
		\State ${\mathcal{P}_l}^{(t)} \gets {\mathcal{P}_l}^{(t)}\setminus \{m\}$
		\State ${\mathcal{P}_m}^{(t)}\gets {\mathcal{P}_m}^{(t)}\setminus \{l\}$
		\EndFor
		\EndFor
		\Until{${\mu}^{(t)}= {\mu}^{(t-1)} $}
	\end{algorithmic}
	
\end{algorithm} 

\subsubsection{Preference Profiles of Players}
\label{subsubsec:PPP}
In this subsection, we first formulate pair selection as a two-sided matching game. In our model, there are two types of cellular users. The first type is eMBB users $\mathcal{E}$, while the second type is URLLC users $\mathcal{U}$. In the two-sided matching game, each player of one side must rank the players of the other side in descending order of priority, which is represented by a preference profile. 

The willingness of URLLC users to adopt the superposition scheme is determined by the BS through the geographical locations of URLLC users. Each eMBB user $e$ ranks the potential URLLC users on the basis of their willingness to adopt the superposition scheme. For a URLLC user, the preference of the eMBB user is high for URLCC users that posses low channel gain so that it can help adopt the superposition scheme according to \eqref{supEmbb} and \eqref{supUrllc}. The URLLC user $u$ creates a preference profile based on the following preference function:  

\begin{equation}
\label{eq:pref_profile_sub}
\mathcal{P}_u = g_{ej}, \forall u \in  \mathcal{U}.
\end{equation} 

Similarly, the preference of the eMBB user is to select a URLLC user whose willingness to adopt the superposition scheme is the highest. Then, in such case, the superposition scheme is selected for the URLLC transmission. The preference profile for the URLLC user $e$ is based on the classification of the superposition according to Definition \ref{def1} expressed as follows:  

\begin{equation}
\label{eq:pref_profile_provider}
\mathcal{P}_e = \theta_n, \forall e \in  \mathcal{E}.
\end{equation}    

Note that once both sides build their respective preference profiles, then, the all eMBB users propose to find their best suited URLLC user based on their preference profiles. We can adopt the deferred acceptance algorithm to execute this process. 

Next, we present our URLLC-eMBB matching algorithm which is based on the deferred acceptance algorithm. In the initialization phase (lines 1~3), all variables are initialized. Then, each eMBB user will send proposals to their most preferred URLLC user based on their preference profiles ${\mathcal{P}_e}^{(t)}$ (line 7). The URLLC users on receiving the proposals compare them with existing proposals and choose the one which is the highest ranked in its preference profile (lines 8~11). Note that if the received proposal is ranked lower in the URLLC profile it is immediately rejected which prevents blocking pairs in our scheme (lines 13~14). These rejected proposals are then removed from both preference lists of emBB and URLLC users (lines 15~19). Note that this process of removal ensures the stability of the matching game and prevents blocking pairs to occur. Finally, we receive a stable matching after a number of iterations once the set of matching pairs to do not change (lines 21).            

Note that the one-to-one matching is used to select the suitable eMBB and URLLC pairs for the superposition. However, in some cases superposition scheme is not possible to adopt due to the channel gains of matched pairs. In such cases, the puncturing scheme is adopted by the URLLC users. 

\subsection{Contract feasibility and optimality}
Using matching theory, which is discussed in previous section, the suitable contracts for URLLC users are designed. To provide the conditions for the feasibility of the designed contracts, the following conditions apply: 

\begin{mydef} \textbf{(Individual Rationality (I.R))}
	For any $u \in \mathcal{U}$, $U_u > 0$. 
\end{mydef}

\begin{mydef} \textbf{(Incentive Compatibility (I.C))}
	For any $u \in \mathcal{U}$, $U_u(\theta_n) \geq U_u(\theta_{n^{'}})$. 
\end{mydef}
In addition to the above conditions, there are sufficient and necessary conditions for the feasibility of the contracts given as follows: 
\begin{mydef} \textbf{(Necessary Condition)}
	For any $u,u^{'} \in \mathcal{U}$, $U_u > U_{u^{'}}$ if and only if $y_{uj} > y_{u^{'}j}$. 
	\label{defnec}
\end{mydef}

A similar proof of the Definition \ref{defnec} is provided in \cite{duan2012cooperative}. This definition states that a URLLC user $u$ who is more willing to opt for superposition has higher utility than a user $u^{'}$, who is less willing to select superposition. 

\begin{algorithm}[!t]
	\caption{Contract-based resource allocation algorithm}\label{finAlgo}
	\begin{algorithmic}[1]		
		\State \textbf{Input:} $J$, $U$, $E$, $\mathcal{\varTheta}$, $\gamma$, ${\mu}^{(t)}$
		\For {$u = 1$ \textbf{to} $U$}	
		\State \textbf{Step1:} Contract Design	
		\State Identify class $\theta_n$ of each URLLC user $u$  
		\State Design contract for URLLC user $u$ to superpose with eMBB user $e$ using the pairing ${\mu}^{(t)}$
		\State Check the contract feasibility and optimality for the matching pair ${\mu}^{(t)}$
		\State \textbf{Step2:} 
		\If {Contract is feasible } 
		\State Compute utility $U_u$ according to (\ref{util_u}) for $\theta_n \geq \theta$
		\Else 	
		\State Compute utility $U_u$ according to (\ref{util_u}) for $\theta_n < \theta$	
		\EndIf
		\EndFor
		\State \textbf{Output:} $P^{*}_{jk}$ 
	\end{algorithmic}
	\label{algo}
\end{algorithm}	

\subsection{Contract-based Problem Formulation}
\label{secContr2}
After solving the adoption of superposition or puncturing scheme ${x_{uj}^k}^*$ and ${z_{uj}^k}^*$ for the URLLC users in \ref{util_u}, the rest of the problem is the power allocation to the URLLC users. In this section, we formulate a contract-based optimization problem which is equivalent to (\ref{eq:obj1}). The proof of equivalence is given in Appendix \ref{appProofEqui}. 
\begin{subequations}\label{eq:obj2}
	\begin{align}
	\max_{\boldsymbol{p}} \tag{\ref{eq:obj2}} \quad 
	& \sum\limits_{ j \in \mathcal{J}} U_j,  \\
	\text{s.t.}\quad 
	&  \label{eq:p2const1}  U_u \geq 0 , \quad \forall u \in \mathcal{U}, \quad \textrm{\textbf{(I.R)}}, \\
	&  \label{eq:p2const2}  U_u(\theta_n) \geq U_u(\theta_{n^{'}}) , \quad \forall u \in \mathcal{U}, \quad \textrm{\textbf{(I.C)}}, \\	
	&\label{eq:p2const4} p_{uj} \geq 0 , \quad \forall u \in \mathcal{U}, j \in \mathcal{J}.
	\end{align}
\end{subequations}     

The objective (\ref{eq:obj2}) is to maximize the profit of BS $j$. The constraints (\ref{eq:p2const1}) and (\ref{eq:p2const2}) are individual rationality (IR) and incentive compatibility (IC) constraints, respectively. Problem (\ref{eq:obj2}) is a constrained maximization problem for $P_{uj}$ and the solution can be found at the boundary of constraint (\ref{eq:p2const2}). Therefore, optimal power allocation, $P_{uj}^{*}$, is performed such that the conditions in (\ref{util_u}) are satisfied. Note that, the power allocated to the URLLC user is same for both puncturing and superposition schemes; however, the URLLC user utility is less in the puncturing case as compared to the superposition case because the user is not given the contribution incentive $\varsigma \beta_e$.

Contract-based resource association is performed as described in Algorithm \ref{algo}. The inputs are the total number of BSs $J$, total number of URLLC users $U$, total number of eMBB users $E$, set of classification $\mathcal{\varTheta}$ of the URLLC users' willingness to opt for superposition, $\gamma$, which is the set of SINR levels for the URLLC and eMBB users, and ${\mu}^{(t)}$, which is set of matching pairs. In Algorithm \ref{algo}, the contract is first designed for every URLLC user $u$ after identifying the class $\theta_n$. Next, the feasibility and optimality of the contract are verified such that the URLLC QoS requirements are satisfied. Then, power allocation is performed by selecting the superposition or puncturing scheme based on the designed contract. 

\begin{table}[!t]	
	\centering
	\resizebox{\columnwidth}{!}{%
	\begin{tabular}{|lll|}
		\hline
		\multicolumn{1}{|l}{\textbf{Parameters}}&  & \multicolumn{1}{l|}{\textbf{Values}} \\ \hline
		eMBB TTI          		&:	& $1$~ms \\ \hline
		URLLC TTI         		&:	& $0.125$~ms  \\ \hline
		MBS radius        		&:	& $1000$~m \\ \hline
		Noise $N$         		&:	& -~97.5 dBm \\ \hline
		URLLC packet size ($m$) &:    & $100$~B \\ \hline
		Frequency $f$           &:    & $2$~GHz   \\ \hline
		eMBB power ($P_{ej}$)   &:    & $0.01$~mW \\ \hline
		Bandwidth $W$           &:    & $5$~MHz   \\ \hline
		Path loss (eMBB) 		&:	& $35.3 +  37.6\log(d_{ej})$ \cite{kazmi2018coordinated}   \\ \hline
		Path loss (URLLC) 		&:    & $16.62+37.6\log(d_{uj}))$ \cite{popovski20185g} \\ \hline                                  
	\end{tabular}}
	\caption{Simulation parameters}
	\label{tab1}
\end{table}

\section{Numerical Results}
\label{secSim}

\begin{figure}[!t]
	\centering			
	\includegraphics[width=0.5\linewidth]{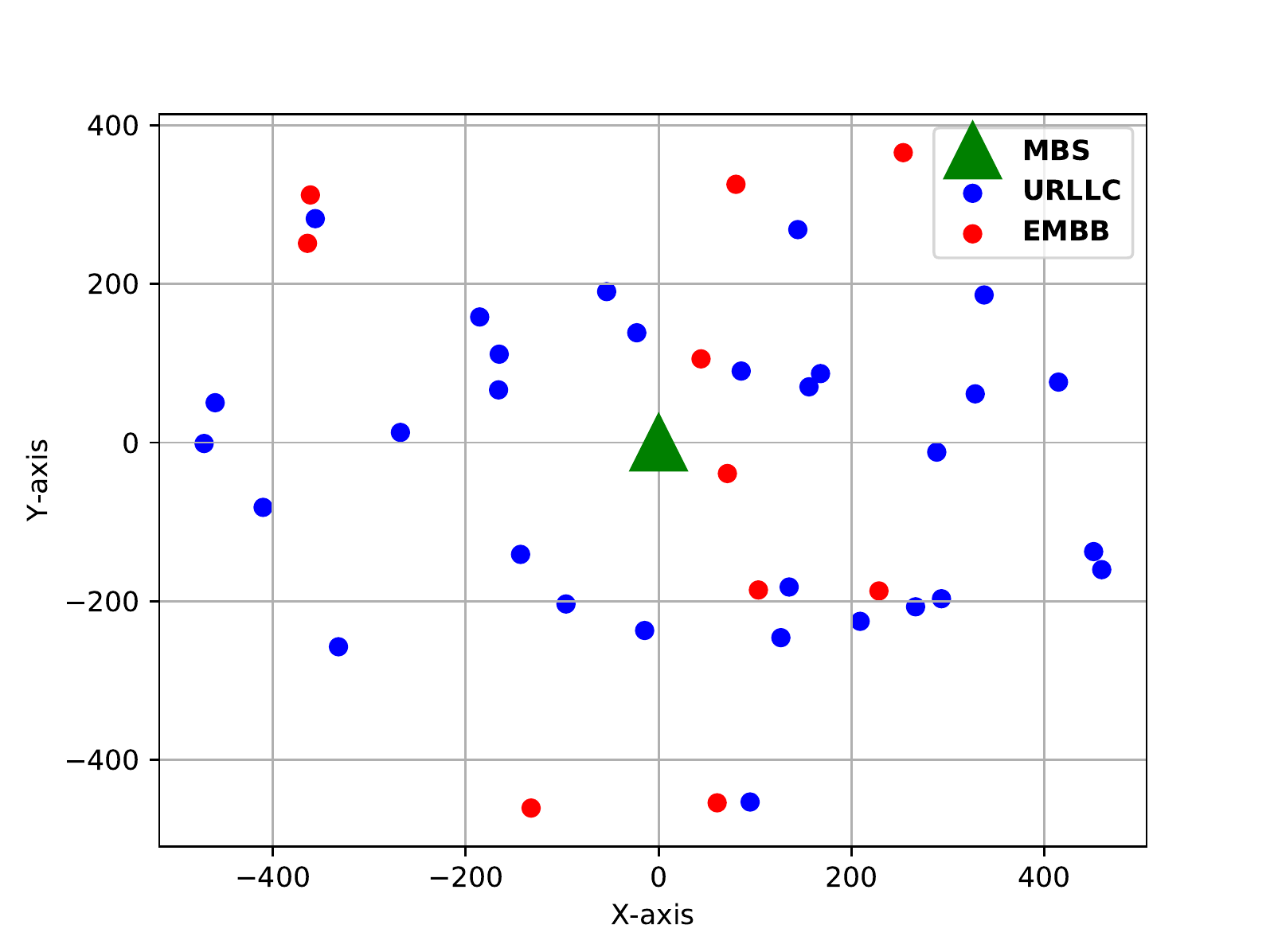}
	\caption{Network topology consisting of a single macro base statoin (MBS) deployed at the center of the network and a number of URLLC and eMBB users uniformly deployed in the area. }
	\label{topo}
\end{figure}

We perform extensive simulations to evaluate the proposed contract-based scheduling scheme in the URLLC and eMBB coexistence network. Firstly, we model a system having a single macro BS (MBS) located at the center of a geographical area of $1000$~m $\times$ $1000$~m, with two coexisting eMBB and URLLC networks. We uniformly deploy URLLC and eMBB users in the area, as illustrated in Figure \ref{topo}. Then, we simulate the network for multiple runs to obtain the average results with a different number of URLLC users. Figure \ref{topo} presents a snapshot of the network topology. Other simulation parameters are summarized in Table \ref{tab1}. We compare the proposed contract-based scheduling scheme with the following two baseline schemes: 1) The \emph{"No URLLC" scheme} refers that there are no URLLC users in the network. This scheme is used to determine the impact of URLLC users and compare the loss in eMBB rate. 2) The \emph{"Puncturing" scheme} refers to the traditional URLLC scheduling scheme in which eMBB traffic is paused during URLLC transmission.

\begin{figure}[!t]
	\centering			
	\includegraphics[width=0.5\linewidth]{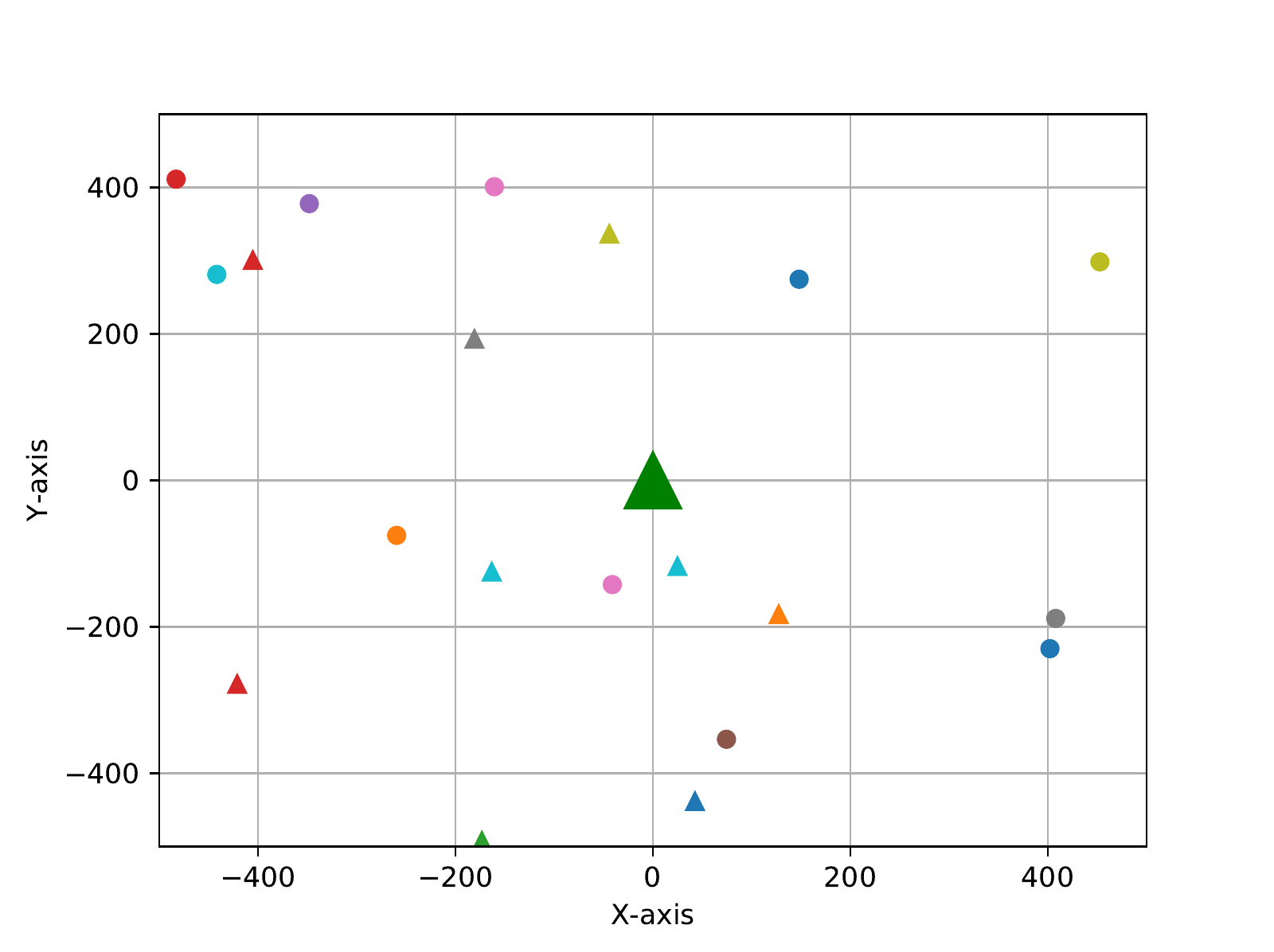}
	\caption{Snapshot of the matching pairs of URLLC and eMBB users. }
	\label{topoMatch}
\end{figure}

Figure \ref{topoMatch} displays the snapshot of the matching pairs of URLLC and eMBB users in the network, where the dots and triangles represent the eMBB and URLLC users, respectively while the same colors are assigned to matched users. 

\begin{figure}[!t]
	\centering			
	\includegraphics[width=0.5\linewidth]{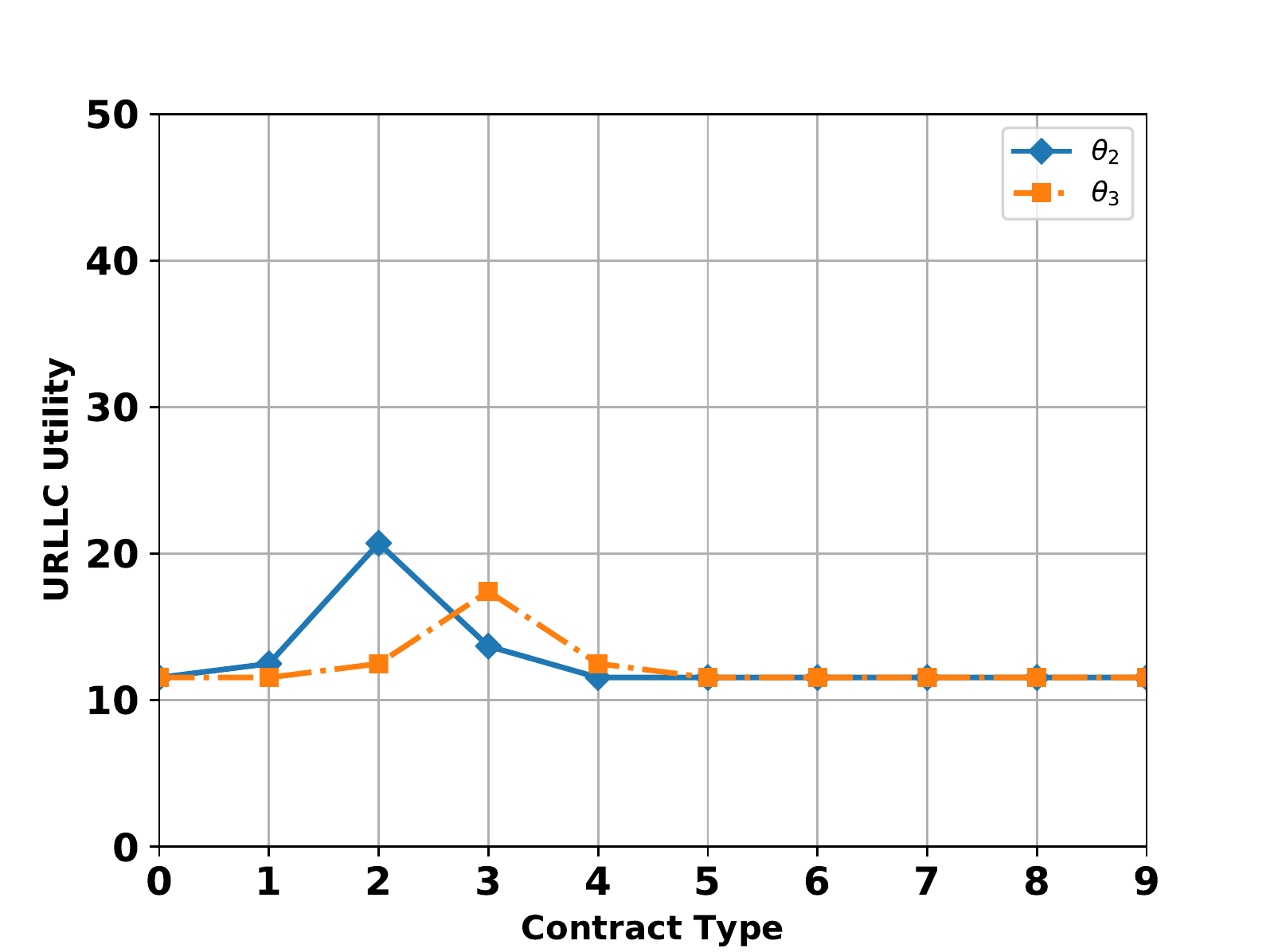}
	\caption{Utility of URLLC users vs. the contract type. }
	\label{contType}
\end{figure}

Figure \ref{contType} shows the dependence of the URLLC utility on the contract type. It can be observed that the user with certain contract type gets the maximum utility as compared to the other users. For instance, the user with type $\theta_2$ will get maximum utility on choosing the contract type $2$.  Therefore each URLLC user chooses the contract according to its type maximize its utility.   

\begin{figure}[t]
	\centering			
	\includegraphics[width=0.5\linewidth]{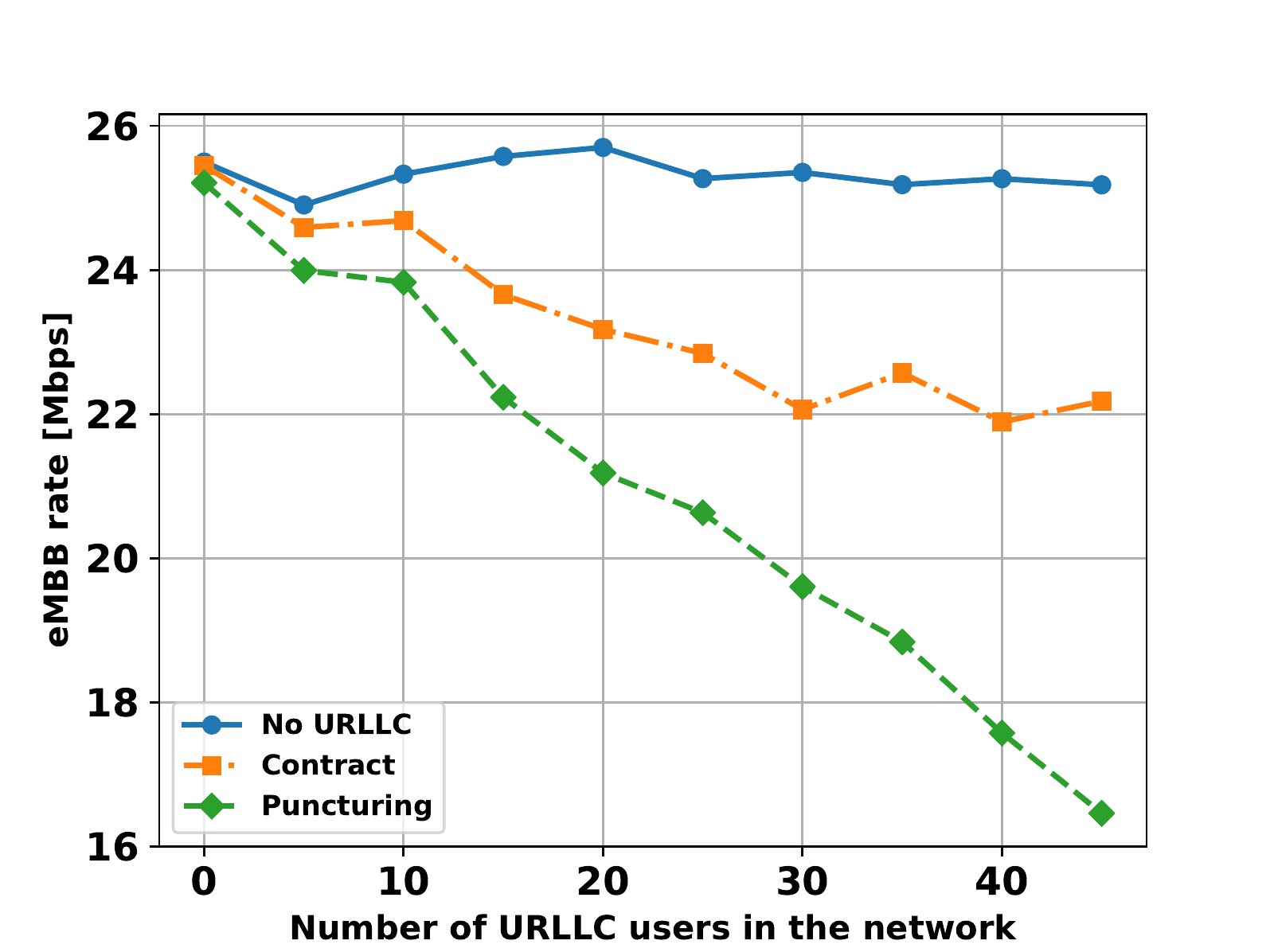}
	\caption{eMBB rate vs. the number of URLLC users in the network.}
	\label{embb}
\end{figure}

\begin{figure}[!t]
	\centering			
	\includegraphics[width=0.5\linewidth]{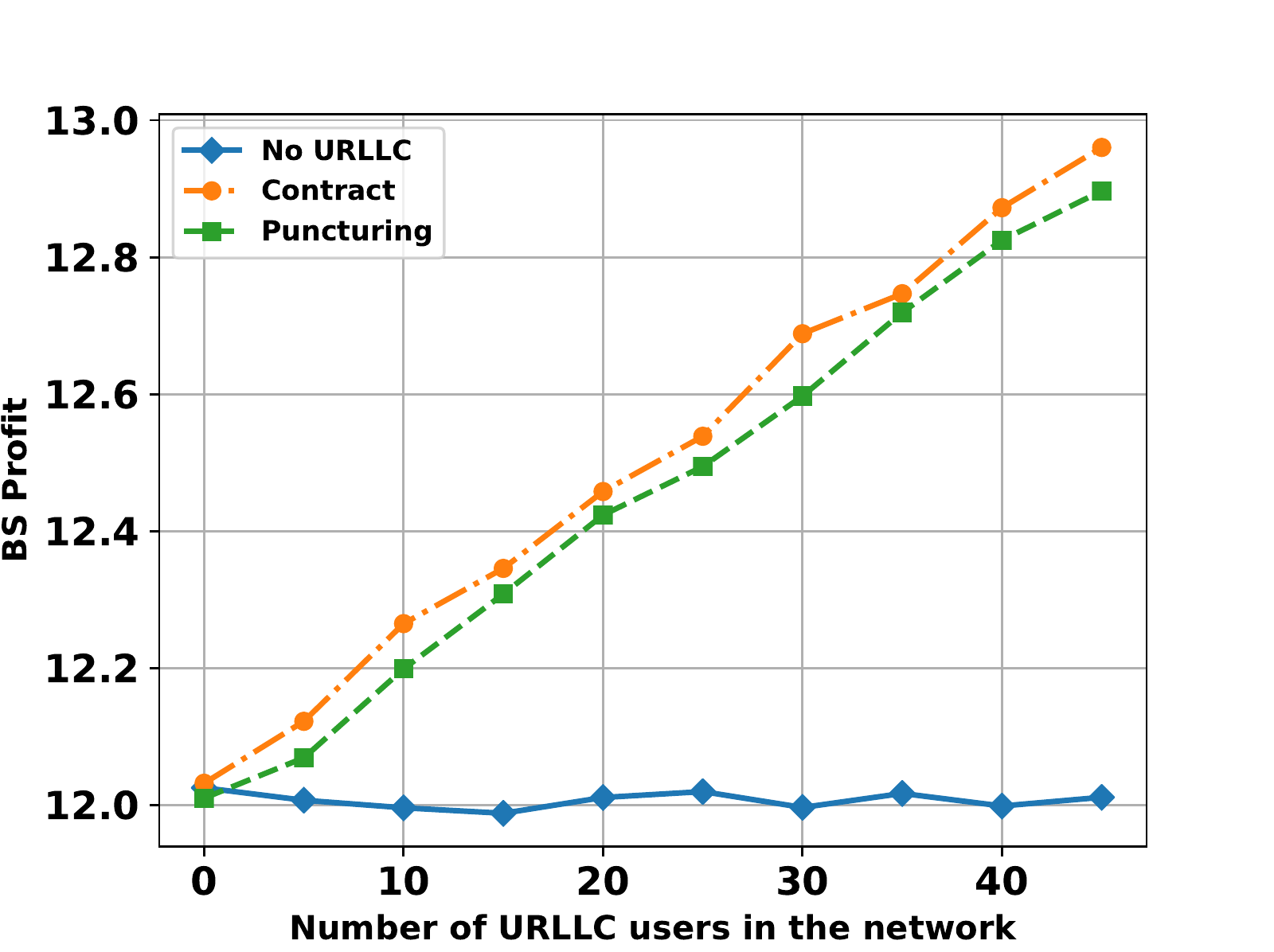}
	\caption{BS profit vs. the number of URLLC users in the network.}
	\label{profit}
\end{figure}

Figure \ref{embb} displays the eMBB rate against an increasing number of URLLC users in the network. It can be observed that as the number of URLLC users increases, the eMBB rate decreases due to scheduling of the URLLC users. It can also be seen that the eMBB rate in the contract-based scheme approaches $63$\% for the non-URLLC case compared to the puncturing scheme. This is due to the use of superposition, in which both URLLC and eMBB users operate on the same channel. 

Figure \ref{profit} illustrates the BS profit against an increasing number of URLLC users in the network. Because the price paid by URLLC users is significantly higher than that paid by eMBB users, the BS profit is increased by serving a larger number of URLLC users. It can be observed that the proposed contract-based scheme provides up to $100.25$\% of the puncturing scheme's profit. This is due to the fact that in the contract-based scheme, the BS offers a portion of the eMBB profit to the participating URLLC users as an incentive.

\begin{figure}[!t]
	\centering			
	\includegraphics[width=0.5\linewidth]{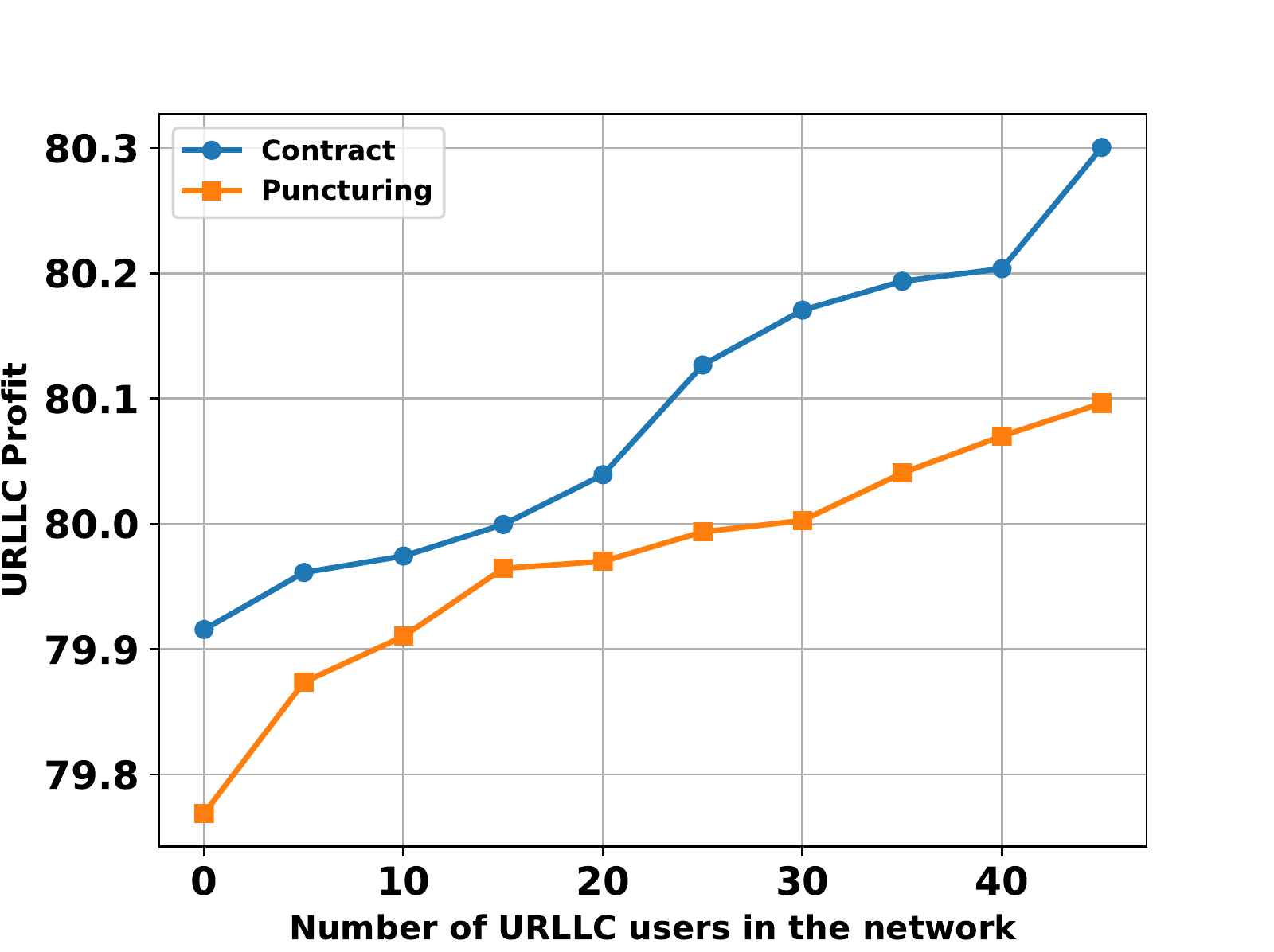}
	\caption{URLLC network profit vs. number of URLLC users in the network.}
	\label{profit_U}
\end{figure}

Figure \ref{profit_U} illustrates the URLLC network utility against an increasing number of URLLC users in the network. Because the proposed contract-based scheme use an incentive mechanism to encourage the URLLC users for the adoption of superposition scheme, the URLLC profit using the proposed scheme is greater than the puncturing scheme. For instance, when there are $30$ number of URLLC users in the network, the proposed contract-based scheme provides up to $106$\% of the No URLLC scheme's profit while the puncturing scheme provides only $103$\% of the profit. The contract-based scheme also offers a portion of the profit to the participating URLLC users as an incentive.

\begin{figure}[!t]
	\centering			
	\includegraphics[width=0.5\linewidth]{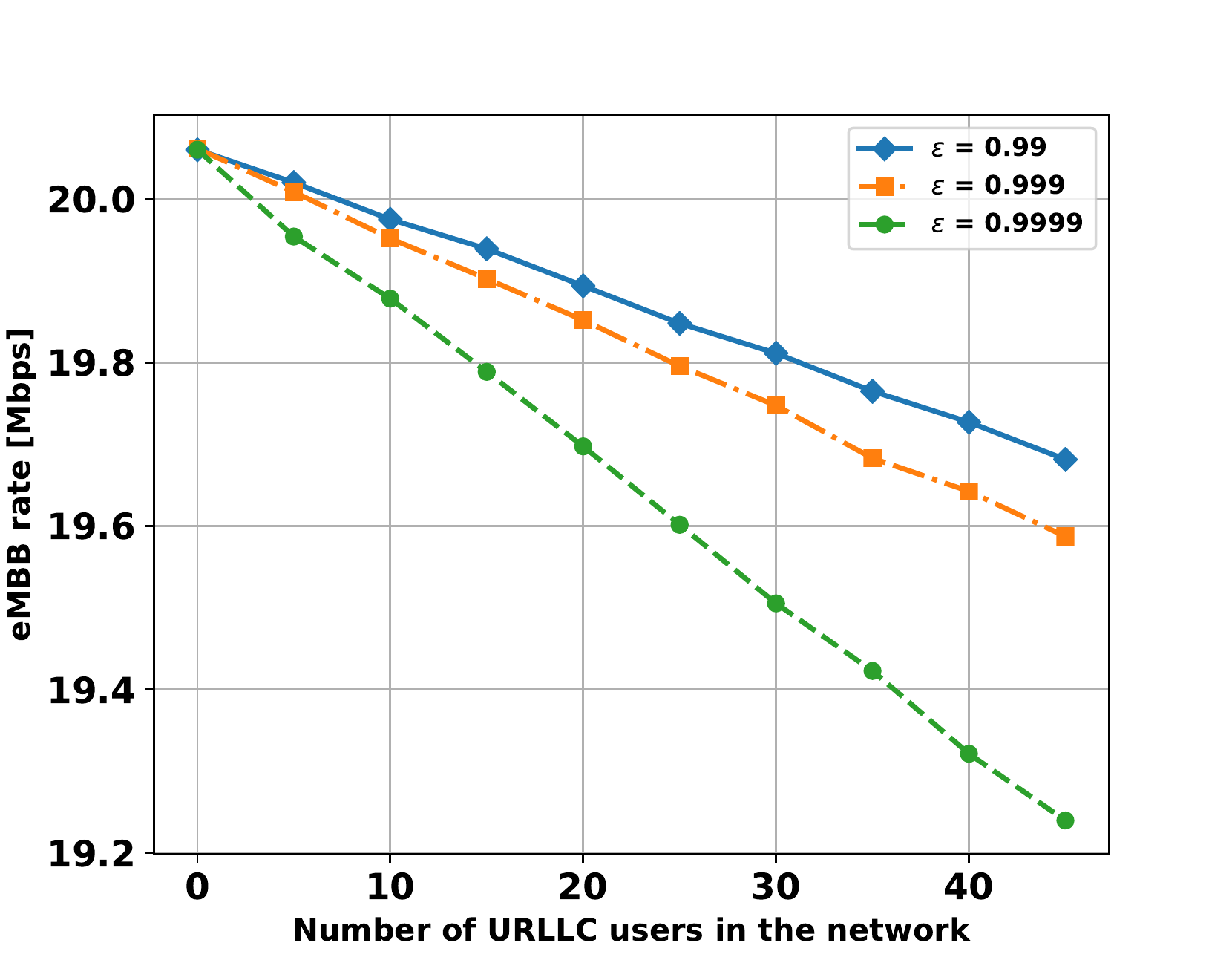}
	\caption{eMBB rate vs. number of URLLC users in the network for different reliability constraints.}
	\label{embb_rel}
\end{figure}

Figure \ref{embb_rel} illustrates the effect of various reliability values on the eMBB rate against an increasing number of URLLC users in the network. We tested the proposed contract-based approach for different values of reliability parameters $\epsilon$. It can be observed that the eMBB rate is reduced for high reliability values. This is due to the fact that the high reliability constraint causes higher SINR requirement for the URLLC users. As a result of this high SINR requirement of SINR requirement, the eMBB users are penalized by choosing the puncturing scheme by the URLLC users to meet the high reliability constraints. Moreover the trends of reduction in eMBB rates with the increase in the number of URLLC users in the network are observed same as shown in Figure. \ref{embb}. 

\section{Conclusion}
\label{secCon}
In this paper, we have addressed the resource allocation problem in URLLC and eMBB coexistence networks. We have formulated an optimization problem to maximize the eMBB network rate with respect to the QoS requirements imposed by the URLLC traffics in the cellular network. Using the puncturing scheme, URLLC users significantly affect the performance of an eMBB network, therefore, we used the contract theory framework to encourage URLLC users to opt for the superposition scheme such that eMBB loss was minimized. This has been achieved by classifying the URLLC users according to their willingness to opt for the superposition scheme, which is favorable for the eMBB network. By using such contracts, resource allocation has been performed to the URLLC users. The numerical results have revealed that the contract-based scheme outperformed the puncturing scheme.

\appendices
\section{Proof of equivalence of problem (9) and problem (14):}
\label{appProofEqui}
The utility of the BS $U_j$ in \eqref{eq:obj2} is a function of the achievable rates of eMBB and URLLC users as given in (\ref{util_bs}). Therefore maximizing $\beta_u$ and $\beta_e$ is equivalent to maximizing the corresponding URLLC and eMBB rate. Moreover, the high prices paid by the URLLC users prioritize their scheduling over eMBB scheduling. From this pricing model, the constraint \eqref{eq:p1const1} is ensured. Further, the individual rationality constraint \eqref{eq:p2const1} ensures the individual utility of each URLLC user to satisfy the reliability requirement given in \eqref{eq:p1const2}. Therefore the problem \eqref{eq:obj1} is equivalent to problem \eqref{eq:obj2}.

%

\bibliographystyle{IEEEtran}
\bibliography{RefURLLC}

\end{document}